\documentclass[12pt]{iopart}
\usepackage{graphicx}
\usepackage{url}
\usepackage{caption}
\usepackage{xcolor}
\usepackage{pdfpages}
\begin{document}

\title {Anomalous thermopower oscillations in graphene-InAs nanowire vertical heterostructures}

\author{Richa Mitra$^1$, Manas Ranjan Sahu$^1$, Aditya Sood$^{2}$, Takashi Taniguchi$^{3}$, Kenji Watanabe$^{3}$, Hadas Shtrikman$^{4}$, Subroto Mukerjee$^{1}$, A.K. Sood$^{1}$, Anindya Das$^{1}$ }

\address{$^1$ Department of Physics, Indian Institute of Science, Bangalore-560012, India.}
\address{$^{2}$ Stanford Institute for Materials and Energy Sciences, SLAC National Accelerator Laboratory, Menlo Park, California, 94025, USA.}
\address{$^{3}$ National Institute for Materials Science, Namiki 1-1, Ibaraki 305-0044, Japan.}
\address{$^{4}$ Department of Physics, Weizmann Institute of Technology, Israel.}
\ead{anindya@iisc.ac.in}
\vspace{10pt}

\begin{abstract}

Thermoelectric measurements have the potential to uncover the density of states of low-dimensional materials. Here, we present the anomalous thermoelectric behaviour of mono-layer graphene-nanowire (NW) heterostructures, showing large oscillations as a function of doping concentration. Our devices consist of InAs NW and graphene vertical heterostructures, which are electrically isolated by thin ($\sim$ 10nm) hexagonal boron nitride (hBN) layers. In contrast to conventional thermoelectric measurements, where a heater is placed on one side of a sample, we use the InAs NW (diameter $\sim 50$ nm) as a local heater placed in the middle of the graphene channel. We measure the thermoelectric voltage induced in graphene due to Joule heating in the NW as a function of temperature (1.5K - 50K) and carrier concentration. The thermoelectric voltage in bilayer graphene (BLG)- NW heterostructures shows sign change around the Dirac point, as predicted by Mott's formula. In contrast, the thermoelectric voltage measured across monolayer graphene (MLG)-NW heterostructures shows anomalous large-amplitude oscillations around the Dirac point, not seen in the Mott response derived from the electrical conductivity measured on the same device. The anomalous oscillations are a signature of the modified density of states in MLG by the electrostatic potential of the NW, which is much weaker in the NW-BLG devices. Thermal calculations of the heterostructure stack show that the temperature gradient is dominant in the graphene region underneath the NW, and thus sensitive to the modified density of states resulting in anomalous oscillations in the thermoelectric voltage. Furthermore, with the application of a magnetic field, we detect modifications in the density of states due to the formation of Landau levels in both MLG and BLG.

\end{abstract}

\section{Introduction}

Over the years, dimensionally mismatched two-dimensional (2D) - one-dimensional (1D) heterostructures \cite{jariwala2017mixed} have demonstrated diverse set of advanced functionalities of the heterojunctions \cite{henning2018charge,jeon2016black,yang2017performance,lee2017mixed}. Recently, these systems have emerged as a fertile ground for realizing novel phenomena like anomalous Coulomb drag \cite{mitra2020anomalous}, and formation of 1D waveguides in 2D materials \cite{cheng2019guiding}. The latter has opened new avenues for applications, where charge carriers in 2D systems are guided through 1D cavities \cite{cheng2019guiding}, holding potential as a method for transmitting information, analogous to photons in optical fibers. Moreover, such engineering of reduced dimensionality offers a strategy to enhance the thermopower (or Seebeck coefficient) of a material \cite{hicks1993thermoelectric,hicks1993effect,dresselhaus1999low,hippalgaonkar2010fabrication}. However, to realize the true potential of these reduced dimensions and manipulate them further, it is important to probe the modulation in the local density of states (DOS), for which non-invasive probes are essential. 
In this regard, non-invasive thermo-electric measurements are suitable tools \cite{Harzheim_2020,cho2013thermoelectric,vera2016direct,zuev2009thermoelectric,park2013atomic,hippalgaonkar2015record}, which can be employed for mixed-dimensional systems. In comparison to quantum capacitance \cite{xia2009measurement,ilani2006measurement} and scanning microscopy techniques \cite{zhang2009origin,stolyarova2007high, cui2017quantized, cui2019thermal, martin2008observation,rutter2007scattering,li2008dirac,
giannazzo2009screening,berweger2015microwave}, thermopower measurements can be easily implemented in 2D-1D systems. 

In conventional thermoelectric measurements \cite{zuev2009thermoelectric,wei2009anomalous,wu2013large}, a heat source is typically placed a few microns away and electrically isolated from the actual device to create a spatially-uniform temperature gradient. In our devices, an InAs nanowire (diameter $\sim 50$nm) placed vertically on the graphene channel (almost in its center) and separated by a thin hexagonal boron nitride spacer (hBN $\sim 10$nm) acts like a local nano-heater. Passing a current through the nanowire (NW) generates heat, and its close proximity leads to heat transfer into the graphene channel and hence a finite thermoelectric voltage. By utilizing this unique heating method, we observe unprecedented large oscillations in the thermoelectric voltage measured across monolayer graphene (MLG) at low temperatures ($\sim$ 1.5K to 20K) as a function of the carrier concentration around the Dirac point. Notably, no oscillations are seen in the Seebeck coefficient calculated using Mott's formula based on the measured resistance. The magnitude of oscillations in the MLG devices reduces with increasing temperature and qualitatively follows the trend of Mott's formula at higher temperatures ($>$ 20K). In contrast, for bilayer graphene (BLG) devices, the thermoelectric voltage does not show oscillations and follows Mott's prediction upto the lowest temperature (1.5K), with an expected sign change around the Dirac point. Notably, our observation differs from the oscillations observed in ref \cite{zuev2009thermoelectric} where the Universal conductance fluctuations (UCF) are manifested as oscillations both in resistance as well as in thermopower at low temperatures.

We explain the observed oscillations of the thermopower with density near the Dirac point, as a consequence of changing effective carrier type inside the graphene channel. We propose that this can arise 
due to the 1D confinement of carriers in graphene, and subsequent formation of sub-bands that modify the local DOS. 
To support this proposal, 
we analyze the carrier density profile in the graphene channel underneath the NW and show that the electrostatic potential of the NW creates a cavity for the charge carriers in graphene, resulting in a modulation of local DOS and hence large oscillations in thermoelectric voltage. To understand why the oscillations are seen only in the thermoelectric response and not in the resistance data, we calculate the temperature profile in the NW-graphene heterostructures. These calculations reveal that the temperature gradient exists predominantly in the region of the graphene that is underneath the NW. The resulting sub-bands from the confining potential give rise to large DOS at certain fillings and hence a high thermoelectric voltage. The absence of such thermoelectric oscillations in NW-BLG heterostructures, due to the weaker confinement, confirms the validity of our physical picture. However, the observed aperiodicity in the thermopower oscillations suggests that the dominant electron-hole puddles near the Dirac point in monolayer graphene may have additional contribution to the oscillations. 
As a step forward, we measure the thermoelectric voltage in the presence of a perpendicular magnetic field and observe periodic oscillations in both the MLG and BLG devices due to the formation of Landau levels (LL). This further confirms our model which suggests that the formation of a confinement potential gives rise to DOS modulation, and subsequent oscillations in thermoelectric voltage in MLG at zero magnetic field.

\section{Experimental details}

\begin{figure*}
\centering
\captionsetup{width=1.0\textwidth}
\includegraphics[width=1\textwidth]{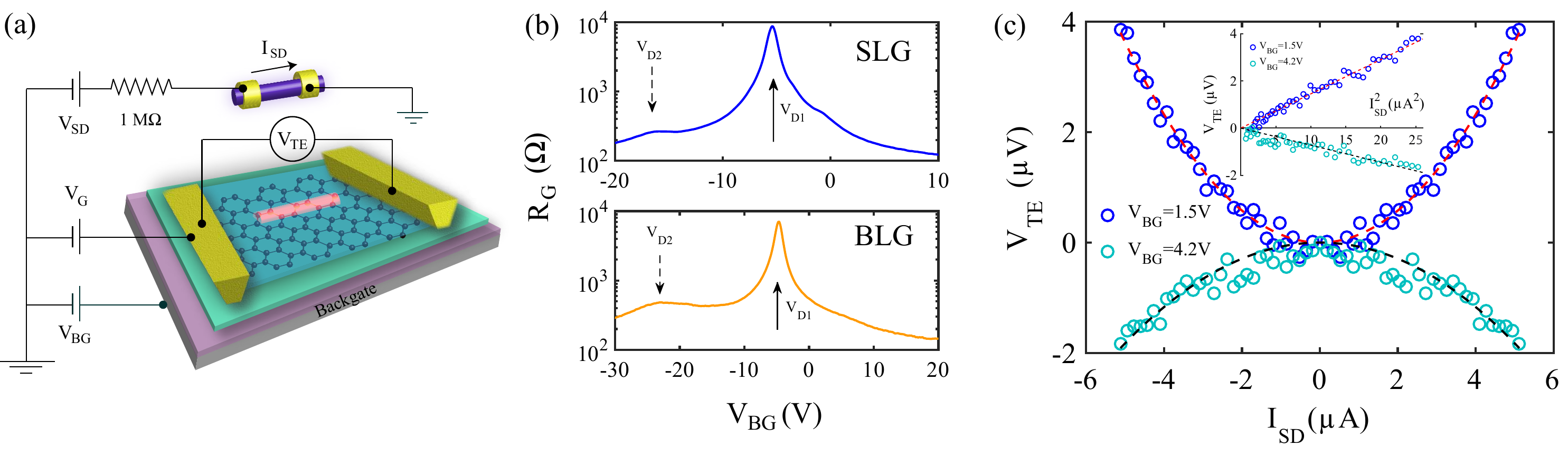}
\caption[width=1.5\textwidth]{(a) Device schematic for measuring thermoelectric voltage in graphene. The heterostructure consists of an InAs NW on top of a  hBN encapsulated graphene stack assembled on Si/Si$O_{2}$ substrate. Thermoelectric voltage ($V_{TE}$) is measured across two probes on graphene while a constant DC current ($I_{SD}$) is passed through the NW. Passing current through the NW heats the region of graphene below the nanowire, and gives rise to a finite thermoelectric voltage in graphene. In this setup carrier density in graphene ($n_{G}$) can be tuned with the backgate voltage ($V_{BG}$). Voltage $V_{G}$ is applied to the graphene to tune the density of the NW. (b) 2-probe resistance of graphene ($R_{G}$) versus backgate voltage for MLG (upper panel) and BLG (lower panel). $V_{D2}$ and $V_{D1}$ indicate (shown by arrows) gate voltages corresponding to charge neutrality points for graphene beneath the NW and for rest of the graphene channel, respectively. (c) $V_{TE}$ plotted with $I_{SD}$ at T=1.5K for two different $V_{BG} = 1.5V$ and $V_{BG} = 4.2V$ indicated by dark blue and sky blue open circles, respectively. The red and black dashed lines are parabolic fits to the data. $V_{TE}$ plotted with $I_{SD}^2$ in the inset. The linear fits (red and black dashed lines) show that $V_{TE}$ has thermoelectric origin.}
\end{figure*}

In this section we describe the device structure of the MLG-NW and BLG-NW devices. The hybrid heterostructures are fabricated by transferring an InAs NW on top of hBN encapsulated MLG or BLG assembled on Si/SiO$_{2}$ substrate using a dry transfer technique \cite{pizzocchero2016hot,wang2013one}. The NW and the graphene are separated by a thin layer of hBN ($\sim 10$nm). The lengths of the NW and graphene channels are $\sim$ 0.4-0.6 $\mu$m and 10-12 $\mu$m, respectively, and the width of the graphene is $\sim$ 10-15 $\mu$m. As shown in Fig. 1a, a constant current $I_{SD}$ passes through the NW when a DC voltage is applied across it through a resistor. Joule heating in the NW creates a temperature gradient in graphene from the position of the NW towards the colder graphene probes. We measure the open circuit voltage ($V_{TE}$) across the graphene channel as shown in Fig. 1a. The carrier concentration in the graphene ($n_{G}$) and NW are tuned by the backgate voltage ($V_{BG}$) and graphene gate voltage ($V_{G}$), respectively.
The 2-probe resistance ($R_{G}$) of MLG and BLG as a function of $V_{BG}$ are shown in upper and lower panels of Fig. 1b, respectively. The two arrows indicate $V_{D2}$ and $V_{D1}$, which are the gate voltages corresponding to charge neutrality points for the graphene that is just beneath the NW, and rest of the graphene channel, respectively; this will be discussed in detail later. The higher value of $\mid{V_{D2}} \mid$ indicates that the graphene region below the NW is more n-doped as compared to the rest of the graphene. Fig. 1c shows the measured open circuit voltage, $V_{TE}$ versus $I_{SD}$ plot for two different gate voltages ($V_{BG}$ = 1.5V and 4.2V) for a MLG device. Both the curves show a quadratic dependence of $V_{TE}$ on $I_{SD}$ as indicated by the dashed lines. The inset showing $V_{TE} \propto I_{SD}^2$ confirms that measured $V_{TE}$ across the graphene channel arises from a thermoelectric response. The positive and negative amplitude of $V_{TE}$ refer to the sign change of the thermoelectric voltage with $V_{BG}$ (Fig. 3). Note that in contrast to our previous work \cite{mitra2020anomalous}, where we concentrated on Coulomb drag (part of the signal that flips with current reversal), here we focus mainly on the non-flipping part of the signal. As discussed in section SI 2, in this measurement, the non-flipping part dominates over the flipping part. 
  
\begin{figure}
\centering
\captionsetup{width=1.0\textwidth}
\includegraphics[width=1.0\textwidth]{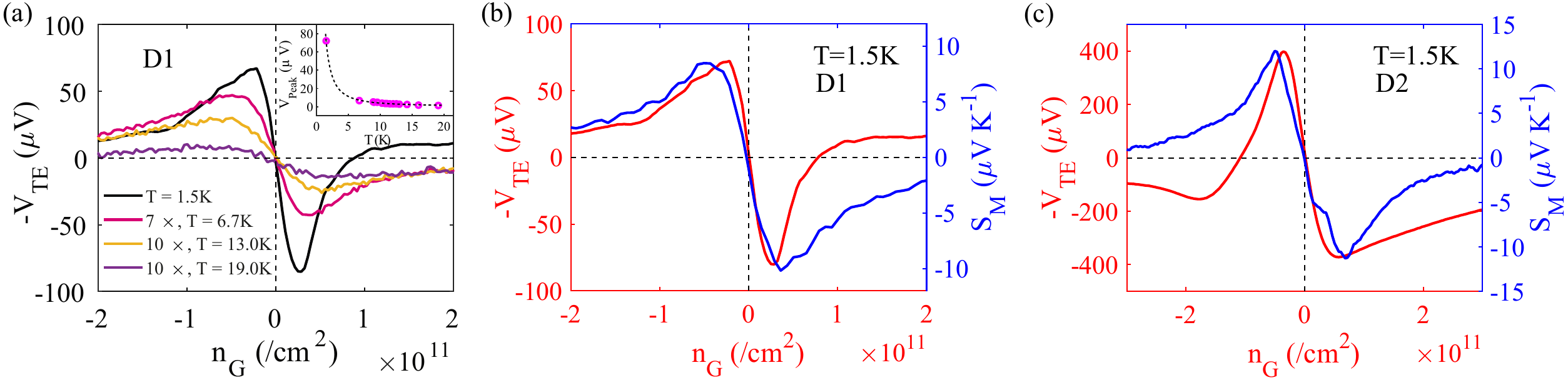}
\caption{ Thermoelectric response of NW-BLG devices: (a) $V_{TE}$ versus $n_{G}$ plotted for different temperatures. The data at 6.7K, 13K and 19K are magnified by 7x, 10x and 10x, respectively. The horizontal and vertical dashed lines indicate the zero voltage and the zero density levels, respectively. $V_{TE}$ flips sign across the charge neutrality point. (Inset) The magnitude of the dip in $V_{TE}$ on the hole side, is plotted versus temperature. The dashed line is a fit to the eye. (b) and (c) are comparison between $V_{TE}$ and Seebeck coefficient for two devices D1 and D2 respectively. The blue lines represent the Seebeck coefficient ($S_{M}$) calculated using the Mott formula (Eqn. 1) at $T=1.5K$ versus $n_{G}$. The red lines are $V_{TE}$ versus $n_{G}$ at T=1.5K for D1 and D2 devices respectively. For BLG, the shape of the $V_{TE}$ matches well with the Seebeck coefficient $S_{M}$. }
\end{figure}

\section{Results}

\begin{figure*}
\centering
\captionsetup{width=1.0\textwidth}
\includegraphics[width=1.0\textwidth]{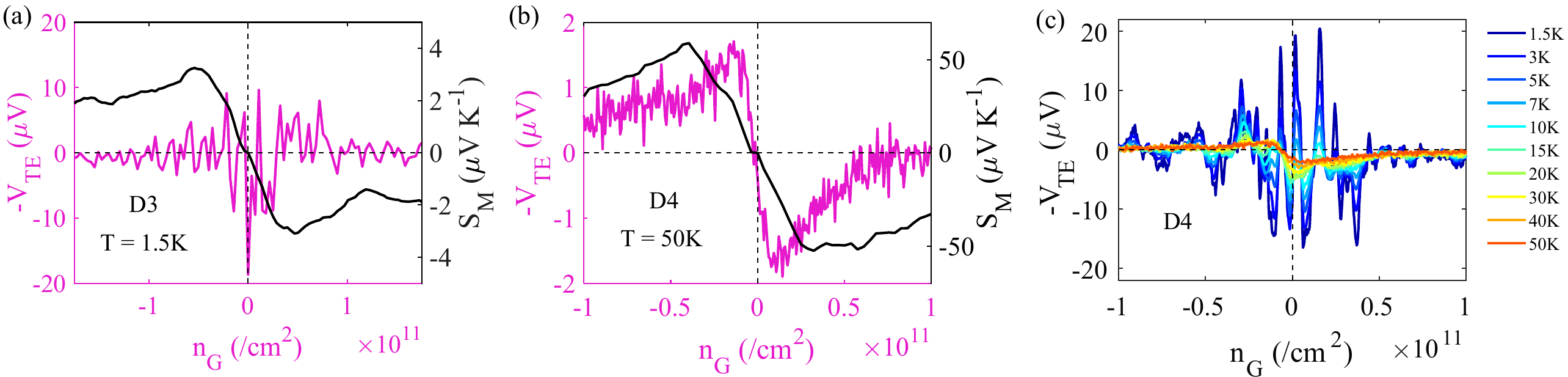}
\caption{ Thermoelectric response of NW-MLG devices: Figure (a) and (b) show comparison between $V_{TE}$ with the Seebeck coefficients for two NW-MLG devices D3 and D4 at T=1.5K and T=50K respectively. The purple lines (left axis) indicate density dependence of $V_{TE}$ for D3 and D4 respectively. The horizontal and vertical dashed lines indicate the zero voltage and Dirac point respectively. On the right axis of (a) and (b) we plot the theoretically estimated Seebeck coefficient $S_{M}$ for D3 and D4 MLG devices (black) based on the measured electrical resistance and the Mott formula. The overall shape of $S_{M}$ matches with $V_{TE}$ at higher temperature, although doesn't match with the $V_{TE}$ at the lower temperature. (c) $V_{TE}$ versus $n_{G}$ plot for device D4 at different temperatures ranging from 1.5K to 50K. In (a) and (c), $V_{TE}$ show rapid oscillations with $n_{G}$ at lower temperatures (T $<$ 20K). The oscillation amplitude reduces as the temperature increases as well as at higher graphene densities.  
}
\end{figure*}
In this section we will first present the thermoelectric response for two BLG devices D1 and D2 followed by for two MLG devices D3 and D4 respectively. Fig. 2a shows measured $V_{TE}$ with carrier density, $n_{G}$, for D1 BLG device at different temperatures ranging from 1.5K to 19K. $V_{TE}$ changes sign across the Dirac point ($n_{G} = 0$). We see from Fig. 2a that $V_{TE}$ decreases in magnitude with increasing temperature. The inset shows the peak magnitude of $V_{TE}$ (for hole side) as a function of temperature, where the dotted black line is a guide to the eye. 
In Fig. 2b and 2c we plot the theoretically estimated Seebeck coefficient $S_{M}$ (blue line) at T =1.5K of the D1 and D2 BLG devices respectively, using the Mott formula:
\begin{equation}
\centering
S_{M} = \frac{\pi^2 k_{B}^2 T}{3e} \frac{\partial ln \sigma}{\partial n} \frac{\partial n}{\partial \mu}
\end{equation}
We utilize the measured resistance (1/$\sigma$) of the BLG device (Fig. 1b lower panel) to evaluate $S_{M}$. We compare the experimentally measured $V_{TE}$ with Mott formula of Seebeck coefficient ($S_{M}$) as shown in Fig. 2b and 2c by plotting $V_{TE}$ versus $n_{G}$ (red line) at the left axis and $S_{M}$ (blue line) at the right axis. 
We find that the overall shape of the $V_{TE}$ for both the NW-BLG heterostructures follows the trend of the Mott formula. 

 
Fig. 3a, 3b present the comparison of experimental $V_{TE}$ with the Seebeck coefficient for two NW-MLG devices D3 and D4 at T=1.5K and T=50K respectively. In the left axis we plot of $V_{TE}$ with graphene density (purple plot), whereas in the right axis we plot the Seebeck coefficient $S_{M}$ versus $n_{G}$ (black line), obtained using Mott's formula for the respective temperatures. Fig. 3c shows the doping dependence of $V_{TE}$ for device D4 at different temperatures ranging from 1.5K to 50K. In Fig. 3a and 3c (for T$<$ 20K), the most striking observation is that the amplitude of 
$V_{TE}$ changes between positive and negative values as the carrier density is varied near the Dirac point. Notably, the periodic nature of the oscillation changes with the graphene density; more periodic near $n_{G} = 0$ and becoming less periodic as we move away from the neutrality point. The oscillations vanish further at higher densities, as well as when the system temperature is increased. For example, at T $ > 20K$, the oscillations disappear and the overall shape of $V_{TE}$ resembles that of the BLG devices (Fig. 2). 
The comparison in Fig. 3a at T=1.5K shows that $S_{M}$ calculated from the device gate response doesn't show any oscillations, otherwise present in the density responses of $V_{TE}$ for the MLG devices. However, the overall shapes of $V_{TE}$ and $S_M$ qualitatively agree at T=50K as shown in Fig. 3b. 
From Fig. 3a and 3c, we find the average period of oscillations in $V_{TE}$ at T=1.5K to be $\delta{n} \sim$ $1.08 \times 10^{10} /cm^2$ and $0.8 \times 10^{10} /cm^2$ respectively which corresponds to energy scale of $\sim 13 meV$ and $10 meV$ ($\delta E_{F} = \hbar v_{F} \sqrt{\pi \delta{n}}$ ) respectively. In Fig. S6 of Supplementary Info, we plot the standard deviation ($SD$) of $V_{TE}$ oscillation amplitude as a function of temperature, where the dashed black line is a guide to the eye. 
The $SD$ is calculated over the density range of $\pm 0.25 \times 10^{11} /cm^{2}$. We see that the $SD$ approaches zero at $\sim$ 20K, which corresponds to a thermal energy broadening ($3.5k_{B}T$) of $\sim 7$ meV ($k_{B}$ being Boltzmann constant $\sim$ 1.38 $\times$ $10^{-23}$ JK$^{-1}$). 
  

\section{Discussion}

In this section, we discuss the origin of thermopower in our unique nano-heating geometry and propose possible scenarios which may give rise to the observed oscillations in $V_{TE}$, otherwise absent in $R_{G}$. In a conventional thermoelectric measurement setup, the heater is usually placed asymmetrically on one side of the sample (few $\mu m$ away) which creates a uniform temperature gradient along the channel length and gives rise to non-zero $V_{TE}$ proportional to the Seebeck coefficient of the material \cite{ashcroft1976solid,goldsmid2010introduction}. In contrast, here, InAs NW placed on top of the graphene serves as a local heater. Since it is placed approximately at the center of the channel, passing a current through the NW creates a temperature profile which peaks at the center of the NW and is expected to decay symmetrically on both side of the NW in graphene. To explain the origin of non-zero $V_{TE}$, we write it in terms of local Seebeck coefficient $S(x)$ and temperature gradient $\frac{\partial T}{\partial x}$ as \cite{Harzheim_2020}:
\begin{equation}
\centering
V_{TE} = \int_{-L/2}^{+L/2} S(x) \bigskip \frac{\partial{T(x)}}{\partial{x}} \bigskip dx
\end{equation} 
where $x$ is the distance from the center of the channel. If we consider the temperature profile to be Gaussian \cite{hu2020enhanced,Harzheim_2020} and centered around the NW, it creates a temperature gradient which is anti-symmetric around the center. For samples with perfect geometrical symmetry i.e. when the NW is placed exactly at the center of the graphene channel, $\frac{\partial T}{\partial x}$ takes equal and opposite magnitudes around the center. If $S(x)$ is uniform or symmetric around the center, we would get zero average $V_{TE}$ from Eqn. (2). It is therefore evident that asymmetry in $S(x)$ or $\frac{\partial T}{\partial x}$ or both can lead to a finite $V_{TE}$. Asymmetry in temperature profile can arise either due to device geometry if the NW is not placed exactly at the center of the graphene channel, or due to asymmetric Joule heating due to different NW contact resistances at the two ends. For $S(x)$, the symmetry can be broken either by the device skewness or due to non-identical density profiles around the center. Most real devices have intrinsic structural asymmetry as shown in Supplementary Information (section SI 1), which can lead to asymmetry in the temperature profile as well as in the Seebeck coefficient. Thus, the finite $V_{TE}$ observed in our NW-BLG heterostructures is not surprising, and $V_{TE}$ will change its sign only once around the Dirac point when the $S(x)$ changes its sign with charge carrier type as seen in Fig. 2.

The non-zero $V_{TE}$ signals in Fig. 2 and 3 are due to the intrinsic asymmetry in the device geometry; this is always present in real samples due to unavoidable uncertainties in device fabrication. Thus, the resultant thermopower \textit{magnitude} is dependent on the inherent asymmetry of the devices, and may vary from device to device. This is reflected as a finite background signal in $V_{TE}$ as shown in Fig. 2b and 2c, where $V_{TE}$ can have positive or negative values at higher $n_{G}$ which is device specific. However, the oscillations in $V_{TE}$ are unaffected by the asymmetry, and are reproducible across multiple devices. 

From the previous discussion, it is clear that to understand the anomalous oscillations of $V_{TE}$ in NW-MLG heterostructures, it is necessary to look beyond Eqn. 2. Since thermopower is directly proportional to $\frac{\partial ln \sigma}{\partial n}$, the changing sign in $V_{TE}$ with the density indicates that the effective carrier type varies as $n$ changes. In order to understand this, we first investigate the local density modulation in graphene. Two different Dirac points in the gate response of $R_{G}$ (Fig. 1b) indicate a non-uniform density profile along the channel. The 2D colormap in Fig. 4b shows how $R_{G}$ evolves with the $V_{BG}$ and $V_{G}$ gate voltages. The black and green dashed lines highlight the variation of the main Dirac point ($V_{D1}$) and the weaker Dirac point ($V_{D2}$) with $V_{BG}$, and from their slopes (section SI 3 for the details) we assign $V_{D2}$ and $V_{D1}$ to the graphene part underneath the NW and to the rest of the graphene channel, respectively. The density mismatch can arise from the trapped charge impurities at the interface of NW- hBN-Graphene hybrid. As shown schematically in Fig. 4a, the density mismatch ($p-n-p$) results in misaligned Fermi energies of the two regions. This can create a cavity for the charge carriers underneath the NW resulting in a modulation of the local DOS as shown by the red line in Fig. 4c (details in the section SI 4). The polarity of $S(x)$ depends on the type of majority charge carrier, and the effective carriers of the cavity will modulate between electron and hole as a function of the Fermi energy shift. Note that $V_{TE} \propto \frac{\partial{\sigma}}{\partial{n}}$ leading to a sign change as the Fermi energy passes through the discrete levels of the cavity as shown in Fig. 4c. However, for our geometry, a symmetric case will produce zero thermoelectric voltage (Eqn. 2). Thus, to get the observed oscillations in $V_{TE}$, the discrete energy levels together with asymmetry in $S(x)$ or $\frac{\partial T}{\partial x}$ are required, the latter being always present in our devices as described in the previous section.

\begin{figure*}
\centering
\captionsetup{width=1.0\textwidth}
\includegraphics[width=0.8\textwidth]{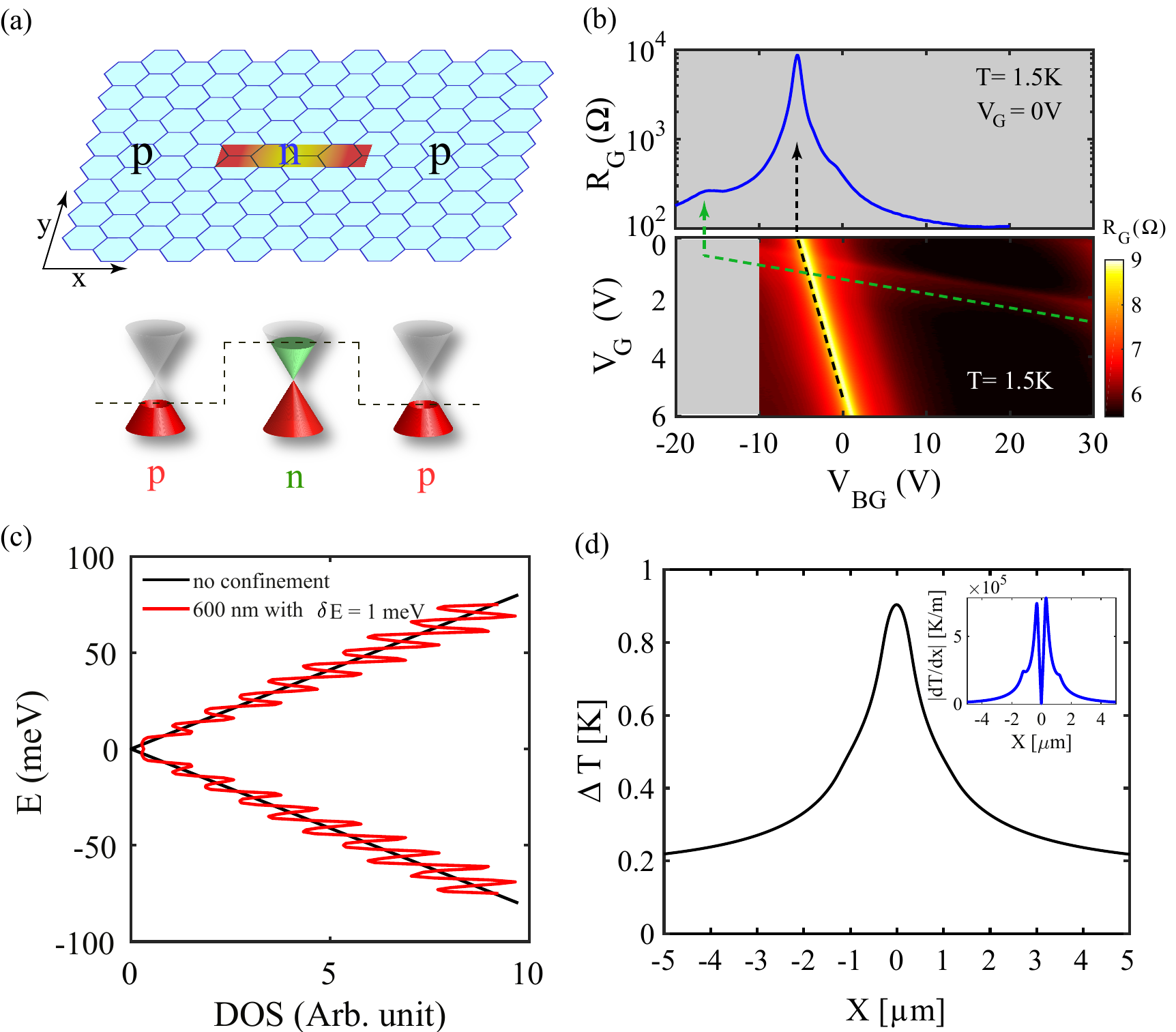}
\caption{\label{fig:epsart} (a) (Upper panel) The hexagonal lattice of graphene is shown in the x-y plane. The red-yellow patch indicates the graphene part underneath the NW. The region of graphene below the NW is n type, while the remaining graphene channel is p type, for the gate voltages between $V_{D2}$ and $V_{D1}$ (Fig. 1b). (Lower panel) The schematic of the band-diagram ($p-n-p$). (b) (Upper panel) Backgate response of $R_{G}$ for MLG plotted in log-scale at T=1.5K. The two peaks indicated by the vertical arrows indicate non-uniform density across the graphene channel. (Lower Panel) 2D colormap of the $R_{G}$ in log-scale plotted as a function of $V_{BG}$ and $V_{G}$ at T=1.5K. The green and black dashed lines highlight the trajectories of the two charge neutrality points with the gate voltages. (c) The red line is the theoretically calculated DOS versus energy for MLG for a cavity with $\sim 600 nm$ width with a Gaussian broadening of $\delta E = 1$meV. The black line corresponds to the case with no confining potential. (d) The total temperature rise $\Delta T$ plotted with position along the $X$ direction. (inset) The absolute value of the temperature gradient $\mid \frac{dT}{dX} \mid$ plotted with position $X$. Details about these thermal calculations are described in SI 7.}
\end{figure*}

To validate our proposed cavity model, we compare the discrete energy levels responsible for the thermoelectric oscillations with the required cavity dimensions. In the previous section we have estimated the energy scale responsible for the $V_{TE}$ oscillations $\sim 7-10 meV$ corresponding to a cavity size of $0.6-0.4 \mu$m, comparable to the length of the NW (details in section SI 5). Note that the other dimension (diameter $\sim$ $50 nm$) will produce discrete energy levels with orders of magnitude larger value, not seen in our experiments. From the values of $V_{D1}$ and $V_{D2}$ (Fig. 1b), the estimated strength of the cavity potential is $\sim 115 meV$ for MLG and $\sim 40 meV$ for the BLG device (details in section SI 5), and thus can explain the absence of oscillations in NW-BLG heterostructure. The weaker confinement is expected in BLG due to large DOS at the Dirac point. To explain the density dependence of the magnitude of $V_{TE}$ oscillations in Fig. 3, we would like to point out the qualitative resemblance between the $V_{TE}$ and Mott's prediction at higher density, where the magnitude of $S(x)$ is expected to decrease with increasing density. Moreover, the effect of screening is also likely to play a role in reducing the strength of the confinement with increasing carrier concentration (SI 5 for the details). Note that the DOS calculation in Fig. 4c does not include the effect of screening, and also we have not considered relativistic effects\cite{zhao2015creating, lee2016imaging, matulis2008quasibound}, which are beyond the scope of this work. 

We now discuss why the oscillations are seen only in the thermoelectric response but not in Mott's prediction derived from the resistance data. We calculate the temperature profile in the NW-graphene heterostructures using a 3D Fourier heat diffusion model (see section SI 7 for details). The temperature profile and its gradient across the graphene are shown in Fig. 4d for uneven Joule heating due to different NW contact resistances at its two ends. Fig. 4d shows that the temperature gradient is dominant in the region of graphene underneath the NW, and thus contributes to the measured $V_{TE}$ significantly according to Eqn. 1. The contribution to $V_{TE}$ from the remaining part of the graphene channel is small as the temperature gradient is close to zero. 
In contrast, the resistance measured across the graphene is dominated by the contribution from the rest of the graphene channel ($\sim 10 \mu m$ X $10 \mu m$) compared to the very small part of graphene ($\sim 50nm$ X $600nm$) just underneath the NW. 

\begin{figure*}
\centering
\captionsetup{width=1.0\textwidth}
\includegraphics[width=1\textwidth]{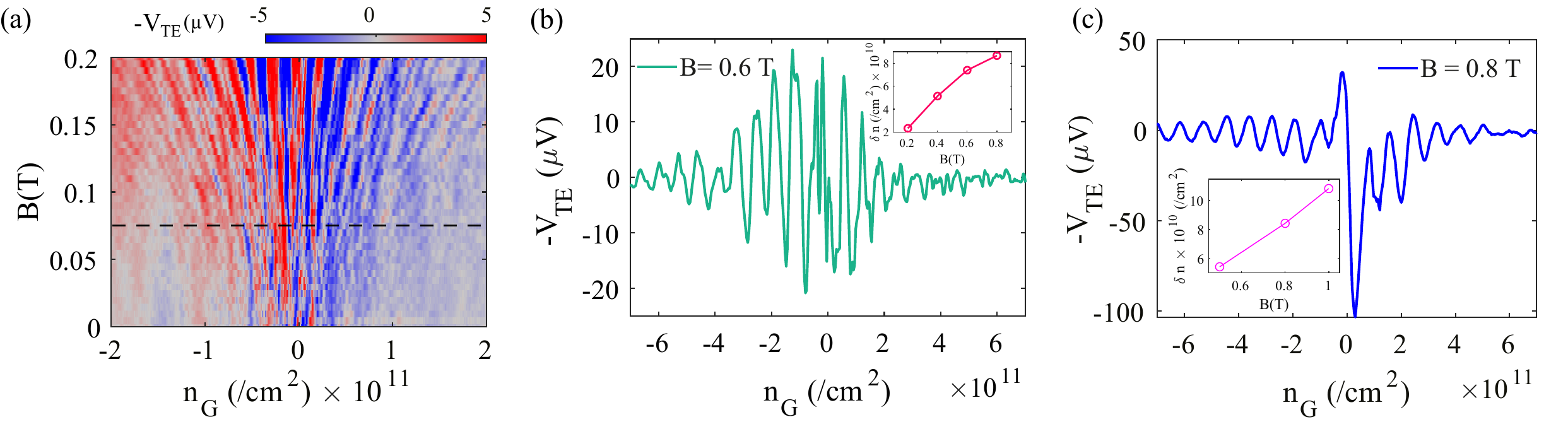}
\caption{ (a) 2D colormap of $V_{TE}$ with graphene density $n_{G}$ and magnetic field from 0 to 0.2T at T=1.5K for the MLG sample. The black horizontal dashed line indicates the onset of Landau level formation captured by the thermopower. $V_{TE}$ versus $n_{G}$ at B = 0.6T for MLG (b), and B= 0.8T for BLG (c). Insets in (b) and (c) show the oscillation period as a function of the magnetic field. The solid lines are guide to the eye.}
\end{figure*}

The NW in this experiment serves a dual purpose, to produce both a highly localized temperature gradient and also a cavity potential in the part of the channel in its immediate vicinity. 
Fig 5a shows a 2D colormap of thermopower with $n_{G}$ and magnetic field for the MLG device, where one can observe LLs for $B \geq 0.1T$. Fig. 5b and 5c show the measured periodic oscillations in $V_{TE}$ for MLG and BLG, respectively, at B $\sim 0.6T$ and at $\sim 0.8T$. As expected, the period of the oscillations increases with increasing magnetic field as shown in the insets of Fig. 5 (SI 6 for details). The period in density $\delta n \sim 0.85\times10^{11}/cm^2$ at B =0.8T for BLG corresponds to $\delta E \sim 3.3meV$ (${\hbar}^2 \pi \delta n / 2m^{*}$), which qualitatively matches the Landau level spacing of $\sim 3.2 meV$ ($\delta E_{LL} \sim \hbar eB/m^{*}$) at B = 0.8T, where $m^{*} \sim 0.03m_{e}$ . For BLG, no oscillations in $V_{TE}$ at B=0 but periodic oscillations at finite magnetic field (section SI 6) further confirm that at zero field there is no modulation in the DOS due to the weaker confinement potential.

Although the concept of cavity formation leading to thermopower oscillation is highly likely, there are other possibilities which may contribute to the oscillations. 
The fact that the oscillations are always observed in the vicinity of the Dirac point and decay quickly as the density is tuned away from $n_{G} = 0$, indicates that the electron-hole puddles can play a role in anomalous oscillations in $V_{TE}$ in MLG. Since the device asymmetry is always intrinsic, the overall effective carrier type changes from electron to hole as the Fermi energy is tuned across the Dirac point. 
The apparent aperiodicity in $V_{TE}$, especially away from the Dirac point suggests that the charge inhomogeneities near the Dirac point may contribute to the oscillations. 

Finally we note that, in this work we compare the measured thermoelectric voltage with the theoretically estimated Seebeck coefficient $S_{M}$ to emphasize that the oscillations are observed only in $V_{TE}$, not in $R_{G}$. This observation excludes the possibility of UCF \cite{zuev2009thermoelectric} as a possible source of oscillations. 


\section{Conclusions}

The thermoelectric response of the NW-MLG devices using InAs NW as a local heater shows anomalous oscillations at low temperatures, which is absent for the NW-BLG devices. The oscillations are only observed in thermopower, not in the electrical resistance. We ascribe them to the changing effective carrier type in the graphene channel with density. By analysing the density profile in graphene, we show that a cavity formed below the nanowire leads to the modification in local density of states which may reflect as alternating sign in the thermoelectric voltage. We also argue that the ubiquitous charge inhomogeneities in graphene near the Dirac point combined with the intrinsic asymmetry may also contribute to the oscillations.  
Thus, our work will pave the way for designing thermoelectric devices using dimensionally mismatched systems, with the potential to enhance the thermopower in two-dimensional materials. We envisage that decoration of graphene by nanostructures resulting in locally enhanced DOS can lead to a high Seebeck coefficient.

\section{Acknowledgement}
AD thanks the Department of Science and Technology (DST), India for financial support (DSTO-2051) and acknowledges the Swarnajayanti Fellowship of the DST/SJF/PSA-03/2018-19. AKS thanks DST for financial support under the Nanomission Project and also thanks DST for the support under the  Year of Science Professorship. KW and TT acknowledge support from the Elemental Strategy Initiative conducted by the MEXT, Japan and the CREST (JPMJCR15F3), JST. We acknowledge Michael Fourmansky for his professional assistance in NWs MBE growth. HS acknowledges partial funding by Israeli Science Foundation (Grants No. 532/12 and No. 3-6799), BSF Grant No. 2014098 and IMOS-Tashtiot Grant No. 0321-4801. HS is an incumbent of the Henry and Gertrude F. Rothschild Research Fellow Chair.

\hspace{10mm}
 
\section*{References}

\bibliographystyle{vancouver}

\bibliography{manuscript}

\newpage
\thispagestyle{empty}
\mbox{}
\includepdf[pages=-]{Supplementary_Info.pdf}

\end{document}